\documentclass[12pt]{article} 

\usepackage{amsfonts,amsmath,amssymb,mathrsfs,verbatim,amsthm}

\begin{document}

\title
{\bf Superconformal indices for  ${\mathcal N}=1$ \\
theories with multiple duals }

\author{V. P. Spiridonov$^1$ and G. S. Vartanov$^{1,2}$
}

\date{}

\maketitle

\vspace{-0.8cm}

\begin{center}
$^1$ Bogoliubov Laboratory of Theoretical Physics, JINR, \\
Dubna, Moscow region 141980, Russia \\
$^2$ University Center,  JINR, Dubna, Moscow region 141980, Russia. \\
E-mails: spiridon@theor.jinr.ru, vartanov@theor.jinr.ru
\end{center}

\begin{abstract}
Following a recent work of Dolan and Osborn,
we consider superconformal indices of four dimensional
${\mathcal N}=1$ supersymmetric field theories related by an electric-magnetic
duality with the $SP(2N)$ gauge group and {\em fixed rank} flavour groups.
For the $SP(2)$ (or $SU(2)$) case with $8$ flavours, the electric theory has
index described by an elliptic analogue of the Gauss
hypergeometric function constructed earlier by the first author.
Using the $E_7$-root system Weyl group transformations for this
function, we build a number of dual magnetic theories.
One of them was originally discovered by Seiberg, the second model was
built by Intriligator and Pouliot, the third one was found by Cs\'aki
et al. We argue that there should be in total 72 theories
dual to each other through the action of the coset group $W(E_7)/S_8$.
For the general $SP(2N),\, N>1,$ gauge group, a similar multiple
duality takes place for slightly more
complicated flavour symmetry groups.
Superconformal indices of the corresponding
theories coincide due to the Rains identity for a
multidimensional elliptic hypergeometric integral
associated with the $BC_N$-root system.
\end{abstract}

\bigskip


\newpage

\tableofcontents

\section{Introduction}

One of the more important recent achievements of mathematical physics consists
of the discovery of elliptic hypergeometric functions -- a new class of
special functions of hypergeometric type (see \cite{Spiridonov1}
for a survey of the corresponding results and relevant literature).
These functions have found applications in the theory of Yang-Baxter
equation, integrable discrete time chains, elliptic Calogero-Sutherland
type models and so on. Although connection with the classical root systems
has been explicitly traced in the structure of many elliptic hypergeometric
functions, their group theoretical interpretation remained largely obscure.

In recent papers R\"omelsberger \cite{Romelsberger} and
Kinney et al \cite{Kinney} have described topological indices
for four dimensional supersymmetric conformal field theories.
As suggested in \cite{Romelsberger}, superconformal indices of
the ${\mathcal N}=1$ models related by Seiberg duality  \cite{S02,Seiberg}
should coincide as a result of some complicated
group theoretical identities. Following R\"omelsberger's ideas,
Dolan and Osborn \cite{DO} have connected superconformal indices
of a number of ${\mathcal N}=1$ supersymmetric field theories with
specific elliptic hypergeometric integrals. Corresponding dual theories
have the same indices due to nontrivial identities for these
integrals \cite{Spiridonov1}.

For example, in  \cite{Spiridonov2} the first author has
discovered the elliptic beta integral opening the door to
a new class of computable integrals. It is described by the following
exact integration formula:
\begin{equation}
\frac{(p;p)_\infty (q;q)_\infty}{2}\int_{\mathbb T}\frac{\prod_{j=1}^6
\Gamma(t_jz^{{\pm 1}};p,q)}{\Gamma(z^{\pm 2};p,q)}\frac{dz}{2\pi i z}
=
\prod_{1\leq j<k\leq6}\Gamma(t_jt_k;p,q),
\label{ell-int}\end{equation}
where six complex parameters $t_j,\; j=1,\ldots,6$,
and two base variables $p$ and $q$ satisfy the inequalities
$|p|, |q|, |t_j|<1$ and the balancing condition
$$
\prod_{j=1}^6 t_j=pq.
$$
Here $\mathbb T$ denotes the unit circle with positive orientation and
$$
\Gamma(z;p,q):=\prod_{j,k=0}^\infty \frac{1-z^{-1}p^{j+1}q^{k+1}}
{1-zp^j q^k}
$$
is the elliptic gamma function.
In (\ref{ell-int}) and below we denote
$(t;q)_\infty:=\prod_{k=0}^\infty(1-tq^k)$ and use the conventions
\begin{eqnarray*}
&&\Gamma(tz^{\pm 1};p,q):=\Gamma(tz;p,q)\Gamma(tz^{-1};p,q),\quad
\Gamma(z^{\pm2};p,q):=\Gamma(z^2;p,q)\Gamma(z^{-2};p,q),\\
&& \Gamma(tz^{\pm 1}w^{\pm 1};p,q):=\Gamma(tzw;p,q)\Gamma(tzw^{-1};p,q)
\Gamma(tz^{-1}w;p,q)\Gamma(tz^{-1}w^{-1};p,q).
\end{eqnarray*}

As shown by Dolan and Osborn \cite{DO}, the left hand side of
formula (\ref{ell-int}) describes the superconformal index of the
``electric" theory with $SU(2)$ gauge group and quark superfields
in the fundamental representation of the $SU(6)$ flavour group.
The ``magnetic" dual theory, suggested by Seiberg in \cite{S02}, does
not have gauge degrees of freedom; the matter sector contains meson superfields
in 15-dimensional antisymmetric $SU(6)$-tensor representation of the
second rank; and its superconformal index is described by the right-hand side of
relation (\ref{ell-int}). This duality provides the simplest
example of the so-called $s$-confining theories.

Seiberg duality is a fundamental concept of the modern quantum
field theory  \cite{S02,Seiberg,Kutasov,KS,IS,I,IP,
Exceptional,Csaki1,Csaki2,Csaki3}. Corresponding models
contain particular sets of fields transforming as
representations of the group $G_{st}\times {G}\times {F}$,
where $G_{st}=SU(2,2|1)$ is the space-time superconformal symmetry
group (containing the $R$-symmetry subgroup $U(1)_R$ rotating supercharges),
${G}$ is the local gauge invariance group, and ${F}$ is the
global flavour symmetry group. Conditionally, electric theories are
considered as manifestations of a unique complicated ``stringy"
dynamics in the weak coupling regime. The magnetic theories are assigned
then to the strong coupling limit. Some of the electric theories
can have more than one dual magnetic partner, as was described
for the first time by Intriligator and Seiberg \cite{IS} (these partners
may differ by symmetries, fields content, or superpotentials).

We have considered systematically superconformal indices of known
${\mathcal N}=1$ supersymmetric theories obeying Seiberg dualities
and compared them with known elliptic hypergeometric integrals.
There are many dualities for ${G}$ composed from
$SU(N),$ $SP(2N),$ $SO(N),$ $G_2$ groups and ${F}$ fixed as products of
$SU(N_f)$ and $U(1)$ groups. For some of them, coincidence of superconformal
indices was established in \cite{DO} as a consequence of previously shown
relations for integrals. As a result of our analysis, we confirm equality of
such indices for several other dual theories and, additionally, we arrive
at  many new conjectures for different elliptic hypergeometric functions
identities. Moreover, from some known integral identities, we arrive at
a good number of new Seiberg dualities.
In this paper we limit ourselves to the models with ${G}=SP(2N)$
and fixed rank flavour groups $SU(8)$ or $SU(8)\times U(1)$
and $SU(6)$ or $SU(6)\times U(1)$, and
their various splits into products of $SU(4), SU(3), SU(2)$, and $U(1)$ groups.
All the dualities described in this paper contain relation (\ref{ell-int})
as a special limiting case of the superconformal index equalities.

The main motivation for us to consider these particular cases of flavour groups
comes from the properties of the elliptic analogue of the Gauss hypergeometric
function constructed by the first author \cite{Spiridonov3,Spiridonov1}.
This function transforms nicely under the action  of the Weyl
group $W(E_7)$ for the exceptional root system $E_7$, and it
is interpreted as a superconformal index for field theories
with ${G}=SP(2)$ (or $SU(2)$) and $F=SU(8)$. Using this fact, we conjecture
existence of distinguished 72 supersymmetric field theories related to each
other by the Seiberg dualities (i.e., all of them should coincide in
the infrared fixed points). The first duality was discovered by Seiberg
himself \cite{Seiberg}. The second dual theory was found by Intriligator
and Pouliot in \cite{IP}. The third admissible magnetic theory was
discovered by Cs\'aki et al in \cite{Csaki2}. Here we argue for the existence
of other models using different interpretation of the
flavour groups. Moreover, our analysis shows that reduction of the number
of flavours from 8 to 6 preserves the multiple duality phenomenon
which indicates on the incompleteness of
the ``$N_f=N_c+1$" Seiberg duality analysis existing in the literature.

For ${G}=SP(2N), N>1$, we use the generalized symmetry transformations for the
type II elliptic hypergeometric integral on the $BC_N$-root system established
by Rains \cite{Rains}. These transformations are described again by the
Weyl group $W(E_7)$. By interpreting the latter integral as a superconformal
index, we conjecture again existence of 72 self-dual theories. Only one of
the corresponding dualities was found earlier in the literature \cite{Csaki1}.
Here we present two new different classes of dualities employing the
antisymmetric tensor matter field.
The 't~Hooft anomaly matching conditions are satisfied for all our dualities
(for smaller flavour groups). The details, as well as a full list of
known dual theories and related superconformal
indices, are described in a separate paper \cite{spi-var:elliptic}.

\section{Superconformal index}

In \cite{Romelsberger} R\"omelsberger has constructed
the superconformal index which counts BPS operators protected
only by one supercharge in four dimensional $\mathcal{N}=1$
superconformal theories. According to his analysis,
first one should determine the index for single particle states which
is given by the formula (for more details on the construction and
the superconformal group, see \cite{Romelsberger,DO})
\begin{eqnarray}\nonumber
&& i(t,x,h,g) = \frac{2t^2 - t(x+x^{-1})}{(1-tx)(1-tx^{-1})}
\chi_{adj}(g)
\\  && \makebox[4em]{}
+ \sum_i
\frac{t^{2r_i}\chi_{R_F,i}(h)\chi_{R_G,i}(g) - t^{2-2r_i}\chi_{{\bar
R}_F,i}(h)\chi_{{\bar R}_G,i}(g)}{(1-tx)(1-tx^{-1})}.
\label{index} \end{eqnarray}
Here the first term represents contribution of gauge fields
belonging to the adjoint representation of the group $G$. The sum $\sum_i$
runs over chiral matter fields $\varphi_i$ transforming as the
gauge group representations $R_{G,i}$ and flavour symmetry group
representations $R_{F,i}$, with  $\chi_{adj}(g)$,
$\chi_{R_F,i}(h)$, and $\chi_{R_G,i}(g)$
being the appropriate characters. Logarithms of the free parameters $t$ and
$x$ play the role of chemical potentials for particular generators of the
superconformal algebra. The terms proportional to $t^{2r_i}$ and
$t^{2-2r_i}$ result from the chiral scalar fields with the $R$-charges $2r_i$ and
fermion descendants with ${\bar \jmath} = {1\over 2}$ of the
conjugate anti-chiral partners whose $R$-charges are equal to $-2r_i$. In order to
determine the index for all gauge singlet operators relevant for
confining theories, formula (\ref{index}) is then inserted into
the ``plethystic" exponential averaged over the gauge group,
which yields the matrix integral
\begin{equation}\label{Ind}
I(t,x,h) \ = \ \int_G d \mu(g) \exp \bigg ( \sum_{n=1}^{\infty}
\frac 1n i \big(t^n ,x^n, h^n , g^ n\big ) \bigg ),
\end{equation}
where $ d \mu(g)$ is the $G$-invariant measure.
Such type of formulas appeared in computation of partition
functions of different statistical mechanics models and quantum
field theories, see, e.g.,
\cite{Sun,Kinney,Naka} and \cite{Plethystic} (where this algorithm
was referred to as the ``plethystic program") or \cite{D2}.

Suppose that we have a chiral superfield with some $U(1)$ symmetry.
Then the corresponding parameter $r$ in the above formula is replaced
by $r+s$, where $s$ is an arbitrary  chemical potential
associated with the generator of $U(1)$. It is convenient to
introduce new variables
$$
p=tx, \quad q=tx^{-1}, \quad z=t^{2s},\quad y=t^{2r}z,
$$
and to assume that $p, q$ are real and $0\leq q,p<1$.
Then the single particle states index takes the form
\begin{equation}\label{1chiral}
i_S(p,q,y) = \frac{t^{2r}z -t^{2-2r} z^{-1}}{(1-tx)(1-tx^{-1})} =
\frac{y - pq/y}{(1-p)(1-q)}.
\end{equation}
As a result of the described index building algorithm,
one obtains the elliptic gamma function \cite{Romelsberger}
\begin{equation}
\Gamma(y;p,q) \ = \ \exp \bigg ( \sum_{n=1}^\infty \frac 1n
i_S(p^n,q^n,y^n ) \bigg ) = \prod_{j,k=0}^{\infty} \frac{1-y^{-1}
p^{j+1}q^{k+1}}{1 - y\, p^j\, q^k}.
\end{equation}
This is precisely how $\Gamma(y;p,q)$ emerged in
the partition function asymptotics for Baxter's eight vertex model \cite{Spiridonov1}.
For the gauge field part one can set
$$
i_V(p,q) \ = \ \frac{2t^2 - t(x+x^{-1})}{(1-tx)(1-tx^{-1})} =
-\frac{p}{1-p}- \frac{q}{1-q} = 1 -
\frac{1-pq}{(1-p)(1-q)}.
$$
Since for $SP(2)$ (or $SU(2)$) gauge group one has $\chi_{adj}(g)=z^2+z^{-2}+1$,
the algorithm yields for different pieces of this character
\begin{eqnarray*}
&& \exp \bigg (
\sum_{n=1}^\infty \frac 1n i_V(p^n, q^n) ( z^{2n} + z^{-2n} ) \bigg )
=\frac{\theta(z^2;p) \theta(z^2;q)}{(1-z^2)^2} \\ \nonumber && \makebox[8em]{}
= \frac{1}{(1-z^2)(1-z^{-2})\Gamma(z^{\pm 2};p,q)},
\\ \nonumber &&  \makebox[2em]{}
\exp \bigg ( \sum_{n=1}^\infty \frac 1n i_V(p^n, q^n)
\bigg ) = (p;p)_{\infty} (q;q)_{\infty},
\end{eqnarray*}
where the theta function is defined as
\begin{equation}\label{theta}
\theta(z;p) \ = (z;p)_\infty(pz^{-1};p)_\infty=
\prod_{j= 0}^\infty (1-zp^j)(1-z^{-1}p^{j+1}).
\end{equation}

\section{Multiple duality for $SP(2)$ gauge group}

\subsection{Electric theory with the flavour group $F=SU(8)$}

In this section we consider multiple duality phenomenon
for a particular electric theory defined as supersymmetric
QCD with the internal symmetry group ${G}\times {F}$, where
$$
{G} \ = \ SP(2),\qquad {F}\ =  SU(8).
$$
All $\mathcal{N}=1$ supersymmetric theories have the
global $R$-symmetry described by $U(1)_R$-group.  So, in the taken version
of SQCD, we have one chiral scalar multiplet $Q$ belonging to the fundamental
representations (denoted as $f$) of $SP(2)$ and $SU(8)$, and the vector
multiplet $V$ in the adjoint representation (denoted as $adj$)
of $SP(2)$ without coupling to $SU(8)$.
We gather information about properties of the fields in Table 1, where
we provide values of $r_i$ for the $U(1)_R$-group in the last column.

\begin{center}
Table 1.
\begin{tabular}{|c|c|c|c|}
  \hline
    & $SP(2)$ & $SU(8)$ & $U(1)_R$ \\  \hline
  $Q$ & $f$ & $f$ & $\frac 14$ \\  \hline
  $V$ & $adj$ & 1 & $\frac 12$ \\  \hline
\end{tabular}
\end{center}

Characters $\chi_R(g)$ for $g \in SP(2)$ are functions of one
complex variable $z$, while the characters $\chi_R(h)$ for $h \in
SU(8)$ are functions of eight complex variables
$$
y=(y_1,y_2,y_3,y_4,y_5,y_6,y_7,y_8), \qquad \prod_{i=1}^8 y_i \ = \ 1.
$$

The single particle state index is given by the expression
\begin{eqnarray}
&& i_E(p,q,z,y) = - \left( \frac{p}{1-p} + \frac{q}{1-q} \right)
\chi_{SP(2),adj}(z) \\  \nonumber && \makebox[2em]{}
+ \frac{1}{(1-p)(1-q)} \left( (pq)^r \chi_{SU(8),f}(y)
\chi_{SP(2),f}(z) - (pq)^{1-r} \chi_{SU(8),\overline{f}}(y)
\chi_{SP(2),\overline{f}}(z)\right),
\end{eqnarray}
where $2r=1/2$ is the $R-$charge of the scalar component of the field $Q$.
The electric index is given then by the following integral
(corresponding characters can be found in the Appendix,
and we borrow the matrix group measures from \cite{DO}):
\begin{eqnarray}
    I_E &=& \frac{(p;p)_{\infty} (q;q)_{\infty}}{2} \int_{{\mathbb T}}
\frac{\prod_{i=1}^8
\Gamma((pq)^{1/4} y_i z^{\pm 1};p,q)}{\Gamma(z^{\pm 2};p,q)}\frac{d z}{2 \pi i z}.
\label{e-index}\end{eqnarray}

In \cite{Spiridonov3} the first author has constructed the
following elliptic hypergeometric function
\begin{equation}
I(t_1, \ldots , t_8;p,q) = \kappa \int_{{\mathbb T}}
\frac{\prod_{j=1}^8 \Gamma(t_jz^{\pm 1};p,q)}{\Gamma(z^{\pm 2};p,q)}
\frac{dz}{z}, \qquad \kappa=\frac{(p;p)_\infty (q;q)_\infty}{4\pi i},
\label{egauss}\end{equation}
with the constraints $|t_j|<1$ for eight complex variables $t_1,
\ldots , t_8 \in \mathbb C$ and the balancing condition
$\prod_{j=1}^8t_j=(pq)^2$.
This integral is interpreted as a natural elliptic analogue
of the Gauss hypergeometric function since it has
many classical properties  \cite{Spiridonov1}.
In particular, it obeys the
following symmetry transformation derived in \cite{Spiridonov3}
(see there formula (6.11) for $n=1$)
\begin{eqnarray}\label{Sp}
&& I(t_1, \ldots , t_8;p,q) = \prod_{1 \leq j < k \leq 4}
\Gamma(t_jt_k;p,q)\Gamma(t_{j+4}t_{k+4};p,q)\, I(s_1, \ldots , s_8;p,q),
\end{eqnarray}
where complex variables $s_j,\, |s_j|<1,$ are connected with $t_j,j=1, \ldots, 8,$
as follows
\begin{eqnarray}
s_j &=& \rho^{-1} t_j, \ j=1,2,3,4, \quad s_j = \rho t_j, \
j=5,6,7,8, \\ \nonumber \rho &=&
\sqrt{\frac{t_1t_2t_3t_4}{pq}}=\sqrt{\frac{pq}{t_5t_6t_7t_8}}.
\end{eqnarray}
This fundamental relation extends the evident $S_8$-permutational
group of symmetries of the integral in parameters $t_j$ to the Weyl group
$W(E_7)$ of the exceptional root system $E_7$ \cite{Rains}.

Evidently, integral (\ref{egauss})
coincides with the electric superconformal index after appropriate
change of variables.
In the following sections we use formula (\ref{Sp}) as a base for
establishing equalities of superconformal indices in known simplest
Seiberg dual theories, as well as for the discovery of new dualities.

Let $e_i,\, i=1,\ldots,8,$ form an orthonormal basis of the Euclidean
space $\mathbb{R}^8$. Denoting as $\langle x, y\rangle$ the
scalar product in this space, we have $\langle e_i, e_j\rangle=\delta_{ij}.$
The root system $A_7$ consists of the vectors $v=\{e_i-e_j,\, i\neq j\},$
and its Weyl group $S_8$ (of dimension $8!$) is generated
by the reflections
\begin{equation}
x\to R_v(x)=x-\frac{2\langle v, x\rangle}{\langle v, v\rangle}\, v,
\label{refl}\end{equation}
acting in the hyperplane orthogonal to the vector $\sum_{i=1}^8e_i$.
This hyperplane vectors $x=\sum_{i=1}^8x_ie_i\in\mathbb{R}^8$ satisfy
the constraint $\sum_{i=1}^8x_i=0$. Evidently, $R_v(\lambda v)=-\lambda v$
for any $\lambda\in\mathbb{C}$ and $R_v^2=1$.

Consider now the change of variables
$t_j=e^{2\pi i x_j}(pq)^{1/4}$ in integral (\ref{egauss}),
which automatically satisfies the balancing condition.
The transformation of parameters in (\ref{Sp}) corresponds then to the reflection
$R_v(x)$ with respect to the vector $v=\frac12(\sum_{i=1}^4e_i-\sum_{i=5}^8 e_i)$
of the length $\langle v, v\rangle=2$ belonging to the root system $E_7$:
\begin{equation}
x'=(x_1',\ldots,x_8')= (x_1-\delta,\ldots,x_4-\delta,x_5+\delta,
\ldots,x_8+\delta),\quad \delta= \frac12 \sum_{i=1}^4x_i.
\label{e7-gen1}\end{equation}
This is the key reflection generating together with $S_8$ the group $W(E_7)$.

Let us apply now $S_8$-group to the set $\{x'\}=R_v(S_8(x))$.
Clearly, the action of its $S_4\times S_4$-subgroup leads to the
vectors that can be obtained by permutation of
$x_1,\ldots, x_8$ in (\ref{e7-gen1}).
However, if we mix  coordinates of $x'$ from $\sigma_1:=\{x_1',\ldots,x_4'\}$
and $\sigma_2:=\{x_5',\ldots,x_8'\}$,
we arrive at new vectors $x''$. $16\times 8!$ of them are
obtained by permutation by one coordinate from $\sigma_1$ and $\sigma_2$.
$18\times 8!$ new vectors appear from permutation by two coordinates from
$\sigma_1$ and $\sigma_2$ (modulo permutation of $\sigma_1$ and $\sigma_2$
themselves which does not lead to new vectors).

Applying again to the derived set of vectors the key reflection
with respect to $v$, we find a number of new elements of the $W(E_7)$-group orbit.
For instance, we obtain
\begin{equation}
\tilde x=(\tilde x_1,\ldots,\tilde x_8)=
(-x_1+\delta,\ldots,-x_4+\delta,-x_5-\delta, \ldots,-x_8-\delta).
\label{e7-gen2}\end{equation}
Application of the $S_8$-group to these new elements yields another set
of $(16+18)\times 8!$ new vectors. Finally, a third application
of the key $R_v$-reflection yields
one more set of independent $8!$ vectors obtained by coordinate permutations
of ${\tilde x}'=(-x_1,\dots,-x_8)$.
This consideration shows that the dimension of $W(E_7)$ is $72\times 8!$
with the coset $W(E_7)/S_8$ consisting of 72 elements generating
transformations $x\to x'\to x''\to \tilde x\to \ldots$ of the described
above form (including the identity transformation).

\subsection{First class of dualities with $F=SU(4)\times SU(4)\times U(1)_B$}

Using relation (\ref{Sp}), we obtain the first magnetic theory with the
internal symmetry groups
\begin{equation}
{G}\ = SP(2),\qquad {F}= SU(4)_l \times SU(4)_r \times U(1)_B.
\label{F1}\end{equation}
It has two chiral scalar multiplets $q$
and $\widetilde{q}$ belonging to the fundamental representation of
$SP(2)$-group, the gauge field in the  adjoint representation $\widetilde{V}$, and
the singlets $M$ and $\widetilde{M}$ in the antisymmetric tensor representations
of $SU(4)$-group. Properties of the fields are summarized in Table 2.

\begin{center}
Table 2.
\begin{tabular}{|c|c|c|c|c|c|}
  \hline
                & $SP(2)$            & $SU(4)$    & $SU(4)$    & $U(1)_B$ & $U(1)_R$                                    \\  \hline
  $q$             & $f$                & $f$            & 1            & $-1$         & $\frac 14$    \\  \hline
 $\widetilde{q}$& $f$   & 1            &$f$& $1$       & $\frac 14$
\\  \hline
  $M$             & 1              & $T_A$            &    1         & $2$        & $\frac 12$                          \\  \hline
  $\widetilde{M}$             & 1              & 1            &    $T_A$         & $-2$        & $\frac 12$                          \\  \hline
  $\widetilde{V}$             & $adj$              & 1            &    1         & 0        &  $\frac 12$  \\  \hline
\end{tabular}
\end{center}

This theory  was found by Cs\'aki et al in \cite{Csaki2},
where it was listed as the third dual theory for the $SU(2)$
gauge group. It differs from the original $SU(2)$ duality
found by Seiberg \cite{Seiberg}, to be described below.

The single particle states index for this magnetic dual theory is
given by the expression
\begin{eqnarray} && \makebox[-2em]{}
i_M(p,q,z,\widetilde{y},\widehat{y}) = - \left( \frac{p}{1-p} + \frac{q}{1-q} \right)
\chi_{SP(2),adj}(z ) \\
\nonumber && \makebox[0em]{}
+\frac{1}{(1-p)(1-q)} \Bigl( (pq)^{r_q}
\frac 1v \chi_{SU(4),f}(\widetilde{y}) \chi_{SP(2),f}(z) -
(pq)^{1-r_q} v \chi_{SU(4),\overline{f}}(\widetilde{y})
\chi_{SP(2),\overline{f}}(z ) \\
\nonumber   && \makebox[0em]{}
+  (pq)^{r_M} v^2 \chi_{SU(4),T_A}(\widetilde{y}) - (pq)^{1-r_M}
\frac{1}{v^2} \chi_{SU(4),\overline{T}_A}(\widetilde{y})   \\
\nonumber && \makebox[0em]{}
+  (pq)^{r_{{\widetilde{q}}}} v \chi_{SU(4),f}(\widehat{y})
\chi_{SP(2),f}(z ) - (pq)^{1-r_{\widetilde{q}}} \frac 1v
\chi_{SU(4),\overline{f}}(\widehat{y})
\chi_{SP(2),\overline{f}}(z ) \\
\nonumber && \makebox[0em]{}
+ (pq)^{r_{\widetilde{M}}}  \frac{1}{v^2}
\chi_{SU(4),T_A}(\widehat{y}) - (pq)^{1-r_{\widetilde{M}}} v^2
\chi_{SU(4),\overline{T}_A}(\widehat{y}) \Bigr),
\end{eqnarray}
where the values of all $r$'s can be read off from the last column of Table 2.
Arbitrary variable $v$ is associated with the $U(1)_B$-group,
its powers are determined by the baryonic charges of the fields.
The characteristic variables $\widetilde{y}_j$
and $\widehat{y}_j$ of the $SU(4)$-groups satisfy the constraints
$\prod_{j=1}^4\widetilde{y}_j=\prod_{j=1}^4\widehat{y}_j=1$.

In order to compare superconformal indices of the electric and magnetic
theories we need matching of the characteristic variables of two different
flavour groups. We denote
$$
\widetilde{y}_j \ = \ v^{-1} y_j, \qquad \widehat{y}_j \ = \
v y_{j+4},\quad j=1,2,3,4,
$$
and set
$$
v=\sqrt[4]{y_1y_2y_3y_4}, \qquad v^{-1}=\sqrt[4]{y_5y_6y_7y_8}.
$$
Applying now formula (\ref{Ind}), we obtain the superconformal
index for the magnetic theory
\begin{eqnarray}\nonumber
    I_M^{(1)} &=&  \frac{(p;p)_{\infty} (q;q)_{\infty}}{2}
\prod_{1 \leq i < j \leq 4} \Gamma((pq)^{r_{M}} y_i y_j;p,q)
 \prod_{5 \leq i < j \leq 8} \Gamma((pq)^{r_{\widetilde{M}}} y_i y_j;p,q)
    \\  && \times
\int_{\mathbb T}
\frac{\prod_{i=1}^4  \Gamma((pq)^{r_q}v^{-2} y_i z^{\pm 1};p,q)
\prod_{i=5}^8  \Gamma((pq)^{r_{\widetilde{q}}}v^2 y_i z^{\pm 1};p,q)}
{\Gamma(z^{\pm 2};p,q)}  \frac{d z}{2 \pi i z}.
\label{tr1}\end{eqnarray}
Using the key formula (\ref{Sp}), we find $I_E=I_M^{(1)}$.
This is a new confirmation of the equality of superconformal
indices for Seiberg dual theories, additional to the results of \cite{DO}.

\subsection{Second class of dualities with $F=SU(4)\times SU(4)\times U(1)_B$}

This dual model has the same flavour group as in the previous section
and two chiral scalar multiplets $q$ and
$\widetilde{q}$ belonging to the fundamental representation of
$SP(2)$, gauge field in the adjoint representation $\widetilde{V}$, and
a singlet $M$. This is the original Seiberg duality for $SU(2)$
group \cite{Seiberg} (it corresponds also to the first $SU(2)$
dual model in \cite{Csaki2}).
The representation content of the model is summarized in Table 3.

\begin{center}
Table 3.
\begin{tabular}{|c|c|c|c|c|c|}
  \hline
                & $SP(2)$            & $SU(4)$    & $SU(4)$    & $U(1)_B$ & $U(1)_R$                                    \\  \hline
  $q$             & $f$                & $\overline{f}$            & 1            & $1$         & $\frac 14$                  \\  \hline
 $\widetilde{q}$& $f$   & 1            &$\overline{f}$& $-1$       & $\frac 14$ \\  \hline
  $M$             & 1              & $f$            &    $f$         & 0        & $\frac 12$   \\  \hline
  $\widetilde{V}$             & $adj$              & 1            &    1         & 0        &  $\frac 12$   \\  \hline
\end{tabular}
\end{center}
The characteristic variables for the $SU(4)$ subgroups are
chosen in the same way as in the previous case.

The single particle index is
\begin{eqnarray} && \makebox[-2em]{}
i_M(p,q,z,\widetilde{y},\widehat{y}) = - \left( \frac{p}{1-p} +
\frac{q}{1-q} \right)
\chi_{SP(2),adj}(z ) \\
\nonumber && \makebox[0em]{} +\frac{1}{(1-p)(1-q)} \Bigl( (pq)^{r_q}
v \chi_{SU(4),f}(\widetilde{y}) \chi_{SP(2),f}(z) - (pq)^{1-r_q}
\frac 1v \chi_{SU(4),\overline{f}}(\widetilde{y})
\chi_{SP(2),\overline{f}}(z ) \\
\nonumber   && \makebox[0em]{} +  (pq)^{r_{{\widetilde{q}}}} \frac
1v \chi_{SU(4),f}(\widehat{y}) \chi_{SP(2),f}(z ) -
(pq)^{1-r_{\widetilde{q}}} v \chi_{SU(4),\overline{f}}(\widehat{y})
\chi_{SP(2),\overline{f}}(z ) \\
\nonumber && \makebox[0em]{} + (pq)^{r_{M}}
\chi_{SU(4),f}(\widetilde{y}) \chi_{SU(4),f}(\widehat{y}) -
(pq)^{1-r_{M}} \chi_{SU(4),\overline{f}}(\widetilde{y})
\chi_{SU(4),\overline{f}}(\widehat{y}) \Bigr).
\end{eqnarray}
The superconformal index itself in this
magnetic theory is found to be
\begin{eqnarray}\label{Sp_2}
&&   I_M^{(2)} =\frac{ (p;p)_{\infty} (q;q)_{\infty}}{2}
\prod_{i=1}^4\prod_{j=5}^8 \Gamma((pq)^{r_{M}} y_i y_j;p,q)
    \\ \nonumber   &&  \makebox[2em]{}
 \times \int_{\mathbb T}\frac{\prod_{i=1}^4
\Gamma((pq)^{r_q} v^2y_i^{-1} z^{\pm 1};p,q)
\prod_{i=5}^8  \Gamma((pq)^{r_{\widetilde{q}}}
v^{-2}y_i^{-1} z^{\pm 1};p,q)}
{\Gamma(z^{\pm 2};p,q)} \frac{d z}{2 \pi i z}.
\end{eqnarray}
Equality $I_E=I_M^{(2)}$ is a direct consequence of transformation (\ref{Sp}).
Namely, it is necessary to repeat once more this transformation
with the parameters $s_3,s_4,s_5,s_6$ playing the role of $t_1,t_2,t_3,t_4$
and permute appropriately parameters in the result
(see, e.g., \cite{Spiridonov1}). Note that this match of
superconformal indices was obtained also in \cite{DO} as the $N=\tilde N =2$
subcase of the $SU(N)\leftrightarrow SU(\tilde N)$ gauge group duality
(see equality (6.12) there).

\subsection{Third dual picture. Flavor group $SU(8)$}

The third type of dual magnetic theories consists of only one model
which was considered by Intriligator and Pouliot \cite{IP} (it was described
as the second $SU(2)$ dual in \cite{Csaki2}). It has
the following symmetry groups
$$
{G}\ =SP(2),\qquad {F}\ = SU(8)
$$
differing from the previous cases. There is one chiral scalar multiplet $q$
in the fundamental representation of
$SP(2)$ and antifundamental representation $\bar f$
of $SU(8)$, the gauge field in the adjoint representation $\widetilde{V}$, and
one singlet $M$, as described in Table 4.

\begin{center}
Table 4.
\begin{tabular}{|c|c|c|c|}
  \hline
   & $SP(2)$ & $SU(8)$ & $U(1)_R$ \\  \hline
  $q$ & $f$ & $\overline{f}$ & $\frac 14$ \\  \hline
  $M$ & 1 & $T_A$ & $\frac 12$  \\   \hline
  $\widetilde{V}$ & $adj$ & 1 & $\frac 12$ \\  \hline
\end{tabular}
\end{center}

The single state index in this case is
\begin{eqnarray} && \makebox[-2em]{}
i_M(p,q,z,y) = - \left( \frac{p}{1-p} + \frac{q}{1-q} \right)
\chi_{SP(2N),adj}(z ) \\
\nonumber && \makebox[2em]{}  + \frac{1}{(1-p)(1-q)} \Biggl\{
(pq)^{\widetilde{r}} \chi_{SU(8),\overline{f}}(y) \chi_{SP(2N),f}(z
) - (pq)^{1-\widetilde{r}} \chi_{SU(8),f}(y)
\chi_{SP(2N),\overline{f}}(z ) \\
\nonumber && \makebox[2em]{}  + (pq)^{r_{M}} \chi_{SU(8),T_A}(y) -
(pq)^{1-r_{M}} \chi_{SU(8),\overline{T}_A}(y) \Biggr\},
\end{eqnarray}

The magnetic index is easily computed to be given by the integral
\begin{eqnarray}
    I_M^{(3)} =\frac{(p;p)_{\infty} (q;q)_{\infty}}{2}
\prod_{1 \leq i < j \leq 8} \Gamma((pq)^{r_{M}} y_i y_j;p,q)
 \int_{\mathbb T}
\frac{\prod_{i=1}^8  \Gamma((pq)^{\widetilde{r}} y_i^{-1}
z^{\pm 1};p,q)}{\Gamma(z^{\pm 2};p,q)} \frac{d z}{2 \pi i  z}.
\label{I3}\end{eqnarray}

The equality $I_E=I_M^{(3)}$ follows from the already established
relation $I_M^{(1)}=I_M^{(2)}$, which is, in a sense, a third sequential
application of transformation (\ref{Sp}) intertwined with the $S_8$-group
actions (see, e.g., \cite{Spiridonov1}). The derived match of
superconformal indices coincides also with the consideration of $N=\tilde N=1$
case of the $SP(2N)\leftrightarrow SP(2\tilde N)$ gauge group
duality in \cite{DO} (see equality (7.12) there).

\subsection{Discussion of the number of dualities and some puzzles}

We have seen that there are at least four field theories dual to each other,
and whose superconformal indices are connected by the specific Weyl group
transformations for the exceptional root system $E_7$. Such
transformations are determined by the coset $W(E_7)/S_8$
of dimension 72. Logically one would expect therefore bigger
number of dualities than we have exhibited.

Trying to model these additional dualities, we considered
the flavour symmetry group
\begin{equation}
{F}=SU(3)_l \times U(1)_1 \times SU(3)_r \times U(1)_2 \times U(1)_B
\label{F2}\end{equation}
and the gauge theory with the field content fixed in Table 5.

{\small
 \begin{center}
\begin{tabular}{|c|c|c|c|c|c|c|c|}
  \hline
  & $SP(2)$  & $SU(3)$ & $U(1)_1$ &  $SU(3)$ & $U(1)_2$ & $U(1)_B$ & $U(1)_R$ \\  \hline
  $q_1$ & $f$ & 1 & $\frac{3}{2}$ & 1 & $\frac{3}{2}$ & $2$ & $\frac 14$ \\ \hline
  $q_2$ & $f$ & $f$ & $\frac{1}{2}$ & 1  & $-\frac{3}{2}$ & $0$ & $\frac 14$ \\ \hline
  $q_3$ & $f$ & 1 & $\frac{3}{2}$ &  1 & $\frac{3}{2}$ & $-2$ & $\frac 14$ \\ \hline
  $q_4$ & $f$ & 1 & $-\frac{3}{2}$ & $f$ & $\frac{1}{2}$ & $ 0$ & $\frac 14$ \\ \hline
  $X_1$ & 1 &  1 & $0$ & $\overline{f}$ & $-2$ & $-2$ & $\frac 12$  \\ \hline
  $X_2$ & 1 &  $\overline{f}$ & $-2$ &  1 & 0 & 2 & $\frac 12$  \\ \hline
  $M_2$ & 1 & $\overline{T}_A$ & $-1$ & 1 & $3$ & $0$   & $\frac 12$  \\ \hline
  $M_4$ & 1 & 1 & $3$ & $\overline{T}_A$ & $-1$ & $0$   & $\frac 12$  \\ \hline
  $\tilde V$ & $adj$ & 1 & 0 & 1 & 0 & 0 & $\frac 12$ \\ \hline
\end{tabular}
\\
Table 5.  Additional dualities of the first class.
\end{center}
}

It is possible to build the superconformal index for this model and find that
it matches with the second class index (\ref{tr1}). However, as it was pointed
to us by A. Khmelnitsky, here one actually has a theory with the flavour group
$F'=SU(4)_l'\times SU(4)_r'\times U(1)_B'$. Let us take the dual
theory of Sect. 3.2 with the flavour group $F'$ and consider decomposition of
the corresponding fields with respect to the subgroup
$SU(3)_l\times U(1)_1' \times SU(3)_r \times U(1)_2'\times U(1)_B'\subset F'$
(evidently, there are more than one such subgroup).
Using the fact that for $SU(3)$ group the $T_A$ and $\bar f$ representations are
Hodge equivalent, one obtains the theory described in Table 5, provided
hypercharges of the corresponding $U(1)$ groups are identified as follows:
\begin{eqnarray*}
&& \makebox[6em]{}
Q_B'=\frac{1}{2}(Q_B+Q_2-Q_1),
\\  &&
Q_1'=-\frac{1}{12} Q_1+\frac{1}{4}(Q_B-Q_2),\quad
Q_2'=-\frac{1}{12} Q_2-\frac{1}{4}(Q_B+Q_1).
\end{eqnarray*}

Similarly one can consider a dual theory with the field content fixed in
Table 6, belonging to the second class of dualities since its superconformal index
matches with (\ref{Sp_2}).

{\small
\begin{center}
\begin{tabular}{|c|c|c|c|c|c|c|c|}
  \hline
   & $SP(2)$ & $SU(3)$ & $U(1)_1$ & $SU(3)$ & $U(1)_2$ & $U(1)_B$ & $U(1)_R$ \\  \hline
  $q_1$ & $f$ & 1 & $-\frac{3}{2}$ & 1 & $-\frac{3}{2}$ & $-2$ & $\frac 14$ \\ \hline
  $q_2$ & $f$ & $\overline{f}$ & $-\frac{1}{2}$ & 1 & $\frac{3}{2}$ & $0$ & $\frac 14$ \\ \hline
  $q_3$ & $f$ & 1 & $-\frac{3}{2}$ & 1 & $-\frac{3}{2}$ & $2$ & $\frac 14$ \\ \hline
  $q_4$ & $f$ & 1 & $\frac{3}{2}$ & $\overline{f}$ & $-\frac{1}{2}$ & $0$ & $\frac 14$ \\ \hline
  $X_1$ & 1 & $f$ & $-1$ & $f$ & $-1$ & 0 & $\frac 12$  \\ \hline
  $X_2$ & 1 & 1 & $3$ & 1 & $3$ & 0 & $\frac 12$  \\ \hline
  $Y_1$ & 1 & $f$ & $2$ & 1 & 0 & $2$ & $\frac 12$  \\ \hline
  $Y_2$ & 1 & 1 & 0 & $f$ & $2$ & $-2$ & $\frac 12$  \\ \hline
  $\tilde V$ & $adj$ & 1 & 0 & 1 & 0 & 0 & $\frac 12$  \\ \hline
\end{tabular}
\\   Table 6. Additional dualities of the second class.
\end{center}
}

Again, one can embed this model into the theory with $F'$ flavour group
with the same relation between $U(1)$-charges as above.
One could claim that the theories of Tables 5 and 6 do not differ from
models of Sects. 3.2 and 3.3, respectively.
However, they differ by the anomaly matching
conditions. Here it is necessary first to explain how we compare global
anomalies of dual theories. Electric theory and third class dual models have
the same $SU(8)$ flavour group and there are no problems in comparison of anomalies.
However in the models of first and second classes the flavour groups are
$SU(4)_l\times SU(4)_r\times U(1)_B$ which leads to the main puzzle
of these dualities. According to 't Hooft, anomalies of the
global symmetries should match in the ultraviolet (UV) and infrared (IR)
regimes. In the second and third class dual models, which supposedly describe
the same IR dynamics,  we miss a large piece of the $SU(8)$ axial currents
needed for comparison with the UV picture of the electric theory.
Surprisingly, this problem was not
discussed in the literature although in many papers this mismatch
in flavour groups for $SU(2)$ gauge group models was noticed (including the
original Seiberg work \cite{Seiberg}). We have found only one
paper by Leigh and Strassler \cite{LS} with partial discussion of
the dynamics in the presence of such an ``accidental symmetry".

So, in \cite{LS} it is claimed that at the IR fixed point the original Seiberg
dual model has actually full $SU(8)$ flavour group, a part of which is
realized in some non-linear non-perturbative way. In support of this
conjecture, rotations of a pair of quark superfields with mass
terms added to the electric theory was considered. In the dual picture
a superpotential was suggested depending on the parameters of this rotation, indicating
on the existence of continuously many dual theories. However, one bothering
issue with considerations of  \cite{LS} is that the manifest flavour symmetry
group is changing its structure abruptly with vanishing of one of the
superpotential parameters. Second, more important,
no explicit flavour $SU(8)$-transformations of the dual theory
were exhibited, their influence on the whole superpotential (e.g., without
adding mass terms) was not established, and no 't Hooft anomaly matching
conditions were verified for the missing part of the global symmetry currents.
All these puzzles show that understanding of the duality for the $SU(2)$
gauge group, where one has an ``accidentally" large flavour group,
is not satisfactory yet.

Return now to the model of Sect. 3.2 with the flavour group $F$.
It is not difficult to check \cite{Csaki2} that its anomalies match
with the anomalies of electric theory for the subgroup $F\subset SU(8)$.
Similar picture holds evidently for the model of Table 5 since it
is equivalent to a similar model with the flavour group $F'$. However,
if we compare
anomalies of the Cs\'aki et al and Table 5 models, there is a nontrivial
possibility to identify the $U(1)_B$ group
in Table 5 (which differs from $U(1)_B'$) with the $U(1)_B$ in Table 2.
To compare anomalies of these two dual models, we need to decompose
fields in Table 2 with respect to the flavour group of Table 5. After
that it can be checked that the anomalies do match indeed. It looks like
that these two Cs\'aki et al type models are related to
each other by some $SU(8)$ flavour space rotation supporting again the Leigh-Strassler
claim about the presence of this hidden symmetry at the IR fixed point.
However, we cannot describe the explicit form of this rotation.
Similar picture holds for the Seiberg type second class dual models
of Tables 3 and 6.

Moreover, one can consider other subgroups of the group
$SU(4)_l\times SU(4)_r\times U(1)_B$:
\begin{eqnarray*}
& (SU(2) \times SU(2)\times U(1))^2\times U(1)_B, \quad
U(1)^3\times SU(2) \times SU(3)\times U(1)_B, & \quad \\
& U(1)^2\times SU(2)^3\times U(1)_B, \quad
U(1)^4\times SU(2)^2\times U(1)_B, \quad & \\
& U(1)^5\times SU(2)\times U(1)_B, \quad
U(1)^6\times U(1)_B &
\end{eqnarray*}
and verify anomaly matchings for them.
Relying on the structure of the coset space $W(E_7)/S_8$
described in the end of Sect. 3.1, we expect that
there will be 35 theories in both first and second classes
of dualities. A diagonal $SU(8)$ matrix can be split into two $4\times 4$
matrices with different entries (up to permutation of these submatrices)
in $\frac12 \left({8\atop 4}\right)=35$ ways. This qualitative counting
corresponds to the number of ways one
can embed $SU(4)_l\times SU(4)_r\times U(1)_B$ into the $SU(8)$ group.

Therefore we expect that the total number of
theories distinguished in UV and related by the duality is equal to 72.
In order to clarify the situation completely, one has to
build superpotentials differentiating all these models.
Also, one may try to build non-linear chiral models for degrees of
freedom associated with the cosets $SU(8)/(SU(4)\times SU(4)\times U(1))$
such that the full anomaly matching conditions will be restored
pairwise for all 72 models. Discussion of such questions lies beyond the scope
of the present paper.

\section{Multiple duality for higher rank symplectic gauge groups}

\subsection{Electric theory with the flavour group $SU(8)\times U(1)$}

Now we pass to investigation of the general $SP(2N)$ gauge group models.
We describe the same multiple duality phenomenon for
$\mathcal{N}=1$ SQCD electric theory
with the overall internal symmetry group ${G}\times {F}$, where
$$
{G} \ = \ SP(2N),\quad N>1, \qquad {F}\ = SU(8) \times U(1).
$$
This theory has  one chiral scalar multiplet $Q$ belonging to the
fundamental representations of $G$ and $F$, the vector
multiplet $V$ in the adjoint representation, and the
antisymmetric $SP(2N)$-tensor field $X$. The field content
is fixed in Table 7.
\begin{center}
Table 7.
\begin{tabular}{|c|c|c|c|c|}
  \hline
    & $SP(2N)$ & $SU(8)$ & $U(1)$ & $U(1)_R$ \\  \hline
  $Q$ & $f$ & $f$ & $- \frac{N-1}{4}$ & $\frac{1}{4}$ \\  \hline
  $X$ & $T_A$ & 1 & 1 & $0$ \\    \hline
  $V$ & $adj$ & 1 & 0 & $\frac 12$  \\  \hline
\end{tabular}
\end{center}
For $N=1$ the field $X$ is absent and $U(1)$-group is completely decoupled.

This electric theory and its one magnetic dual were considered in
\cite{Csaki1}. However, there are more dualities similar to the $SP(2)$ group case.
The single particle states index is
\begin{eqnarray}
&& i_E(p,q,z,y) = - \left( \frac{p}{1-p} + \frac{q}{1-q} \right)
\chi_{SP(2N),adj}(z)
\\  \nonumber && \makebox[2em]{} +
\frac{1}{(1-p)(1-q)} \Bigl\{ (pq)^{r_X} \chi_{SP(2N),T_A}(z) -
(pq)^{1-r_X} \chi_{SP(2N),\overline{T}_A}(z)
\\  \nonumber && \makebox[2em]{}
+  (pq)^{r_Q}
\chi_{SU(8),f}(y) \chi_{SP(2N),f}(z) - (pq)^{1-r_Q}
\chi_{SU(8),\overline{f}}(y) \chi_{SP(2N),\overline{f}}(z) \Bigr\},
\end{eqnarray}
where characters $\chi_R(g)$ for $g \in SP(2N)$ are functions of free $N$
complex variables $z_j, \ j=1, \ldots, N$. We denote also
$$
r_Q \ = \ R_Q +e_Q s,\qquad r_X \ = \ e_X s,
$$
where $2R_Q=1/2$ is the $R$-charge of the $Q$-field,
$e_Q= - (N-1)/4$ and $e_X=1$ are the $U(1)$-group hypercharges, and
$s$ is an arbitrary chemical potential for the latter abelian group.
The electric superconformal index is then
\begin{eqnarray}\nonumber
    && I_E = \frac{(p;p)_{\infty}^N (q;q)_{\infty}^N }{2^N N!}
\Gamma((pq)^s;p,q)^{N-1}
\int_{{\mathbb T}^N}\prod_{1 \leq i < j \leq N}
\frac{\Gamma((pq)^s z_i^{\pm 1} z_j^{\pm 1};p,q)}
{\Gamma(z_i^{\pm 1} z_j^{\pm 1};p,q) }
    \\     &&   \makebox[4em]{} \times
\prod_{j=1}^N
\frac{\prod_{k=1}^8  \Gamma((pq)^{r_Q} y_k z_j^{\pm 1};p,q)}
{\Gamma(z_j^{\pm 2};p,q)} \frac{d z_j}{2 \pi i z_j}.
\label{IE}\end{eqnarray}
We have the constraint $\prod_{k=1}^8y_k=1$, which coincides
with the balancing condition for this elliptic hypergeometric
integral due to a special choice of  the $R$-charge of the chiral
scalar multiplet $Q$.

Now we construct a number of $SP(2N)$-dual theories from the Rains
symmetry transformation \cite{Rains} for the following
higher rank $BC_N$-root system generalization of integral (\ref{egauss}):
\begin{eqnarray}\label{Rains}
&& I(t_1, \ldots , t_8;t,p,q) = \prod_{1 \leq j < k \leq 8}
\Gamma(t_jt_k;p,q,t) \frac{(p;p)_\infty^N (q;q)_\infty^N}{2^N N!}
\nonumber \\ && \makebox[2em]{} \times
\int_{{\mathbb T}^N} \prod_{1 \leq j < k \leq N}
\frac{\Gamma(tz_j^{\pm 1}z_k^{\pm 1};p,q)}{\Gamma(z_j^{\pm
1}z_k^{\pm 1};p,q)} \prod_{j=1}^N \frac{\prod_{k=1}^8
\Gamma(t_kz_j^{\pm 1};p,q)}{\Gamma(z_j^{\pm 2})} \frac{dz_j}{2\pi iz_j},
\end{eqnarray}
where nine variables $t,t_1,\ldots , t_8 \in {\mathbb C}$ satisfy the
balancing condition
$$
t^{2N-2}\prod_{j=1}^8 t_j \ = \ (pq)^2
$$
and the inequalities $|t|,|t_j|<1$. Here
$$
\Gamma(z;p,q,t) \ = \  \prod_{j,k,l=0}^{\infty}
(1-zt^jp^kq^l)(1-z^{-1}t^{j+1}p^{k+1}q^{l+1})
$$
is the elliptic gamma function of the second order
satisfying the key $t$-difference equation
$$
\Gamma(tz;p,q,t) \ = \ \Gamma(z;p,q) \Gamma(z;p,q,t).
$$
Rains has proved the following $W(E_7)$-group transformation for integrals
(\ref{Rains}):
\begin{equation}\label{Rains1}
I(t_1, \ldots , t_8;t,p,q) \ = \ I(s_1, \ldots , s_8;t,p,q),
\end{equation}
where we denoted the variables
\begin{eqnarray}
&& s_j = \rho^{-1} t_j, \ j=1,2,3,4, \qquad s_j = \rho t_j, \
j=5,6,7,8,
\\ \nonumber && \makebox[2em]{}
\rho =
\sqrt{\frac{t_1t_2t_3t_4}{pqt^{1-N}}}=\sqrt{\frac{pqt^{1-N}}{t_5t_6t_7t_8}}.
\end{eqnarray}
We describe a group theoretical interpretation of integral (\ref{Rains})
and use relation (\ref{Rains1}) for equating superconformal indices
of the dual theories. We conjecture again that there are 72 theories dual
to each other with only 4 of them looking essentially different.

\subsection{First class of dualities}

The first magnetic theory has the symmetry groups
$$
{G}\ =\ SP(2N), \qquad {F}\ =\ SU(4)_l\times SU(4)_r\times U(1)_B \times U(1).
$$
It contains two chiral scalar multiplets $q$ and
$\widetilde{q}$ belonging to the fundamental representations of
$SP(2N)$, gauge field in the adjoint representation $\widetilde{V}$,
the anti-symmetric tensor representation $\widetilde{Y}$,
and the singlets $M_J$  and $\widetilde{M}_J, \ J=0, \ldots , N-1$,
as described in Table 8. Similar to $N=1$ case, we expect that there
are 35 dual models in this class.
\begin{center}
Table 8.
\begin{tabular}{|c|c|c|c|c|c|c|}
  \hline
                & $SP(2N)$            & $SU(4)$    & $SU(4)$    & $U(1)_B$ & $U(1)$ & $U(1)_R$
\\  \hline
  $q$             & $f$                & $f$            & 1            & $-1$   &   $-\frac{N-1}{4}$   &  $\frac{1}{4}$  \\  \hline
 $\widetilde{q}$& $f$   & 1            &$f$&  $1$   &  $-\frac{N-1}{4}$    &  $\frac{1}{4}$ \\  \hline
  $Y$             & $T_A$              & 1            &    1         & 0    &  1    & 0                         \\  \hline
  $M_J  $             & 1         & $T_A$            &    1       & $2$     &  $\frac{2J-N+1}{2}$   & $\frac 12$         \\  \hline
  $\widetilde{M}_J $    & 1      & 1            &    $T_A$       & $-2$    &  $\frac{2J-N+1}{2}$    & $\frac 12$     \\  \hline
  $\widetilde{V}$             & $adj$              & 1            &    1         & 0     &  0    &  $\frac 12$   \\  \hline
\end{tabular}
\end{center}
In this and all other tables given below the capital index $J$ takes the values
$0,\ldots,N-1$, which is not mentioned further for saving space.

The single particle state index is
\begin{eqnarray} && \makebox[-2em]{}
i_M(p,q,z,\widetilde{y},\widehat{y}) = - \left( \frac{p}{1-p} + \frac{q}{1-q} \right)
\chi_{SP(2N),adj}(z ) \\
\nonumber && \makebox[2em]{}  + \frac{1}{(1-p)(1-q)} \Biggl\{
(pq)^{r_Y} \chi_{SP(2N),T_A}(z) - (pq)^{1-r_Y}
\chi_{SP(2N),\overline{T}_A}(z ) \\
\nonumber && \makebox[2em]{} + (pq)^{r_q}\frac{1}{v}
  \chi_{SU(4),f}(\widetilde{y}) \chi_{SP(2N),f}(z) -
(pq)^{1-r_q}v \chi_{SU(4),\overline{f}}(\widetilde{y})
\chi_{SP(2N),\overline{f}}(z ) \\ \nonumber && \makebox[2em]{}
 + (pq)^{r_{{\widetilde{q}}}} v \chi_{SU(4),f}(\widehat{y})
\chi_{SP(2N),f}(z ) - (pq)^{1-r_{\widetilde{q}}}\frac{1}{v}
\chi_{SU(4),\overline{f}}(\widehat{y})
\chi_{SP(2N),\overline{f}}(z ) \\
\nonumber   && \makebox[2em]{}  + \sum_{J=0}^{N-1}
\Bigl( (pq)^{r_{M_J}}v^2
\chi_{SU(4),T_A}(\widetilde{y}) - (pq)^{1-r_{M_J}}\frac{1}{ v^2}
 \chi_{SU(4),\overline{T}_A}(\widetilde{y})   \\
\nonumber && \makebox[4em]{}  +
  (pq)^{r_{\widetilde{M}_J}}\frac{1}{v^2}
 \chi_{SU(4),T_A}(\widehat{y}) - (pq)^{1-r_{\widetilde{M}_J}}v^2
\chi_{SU(4),\overline{T}_A}(\widehat{y}) \Bigr) \Biggr\},
\end{eqnarray}
where
$$
r_q \ = \ R_q - \frac{N-1}{4} s, \ \ \
r_{\widetilde{q}} \ = \ R_{\widetilde{q}} - \frac{N-1}{4} s, \ \ \ r_Y \
= \ s,$$
$$
r_{M_J} \ = \ R_{M_J} - \frac12 (N-1-2J) s, \ \ \
r_{\widetilde{M}_J} \ = \ R_{\widetilde{M}_J} - \frac12 (N-1-2J)s.
$$
For the comparison with the electric theory we denote the characteristic variables
as $v=\sqrt[4]{y_1y_2y_3y_4}$ and
$\widetilde{y}_j \ = \ v^{-1}y_j, \,\widehat{y}_j \ = \ vy_{j+4},\, j=1,2,3,4.$
As a result, we find the superconformal index in this magnetic theory
\begin{eqnarray}  \nonumber &&
    I_M^{(1)} =
\prod_{J=0}^{N-1} \prod_{1 \leq i < j \leq 4}
\Gamma((pq)^{r_{M_J}} y_i y_j;p,q)
\prod_{5 \leq i < j \leq 8} \Gamma((pq)^{r_{\widetilde{M}_J}} y_i y_j;p,q)
    \\  \nonumber   &&  \makebox[2em]{} \times
\Gamma((pq)^{s};p,q)^{N -1}\frac{(p;p)_{\infty}^{N} (q;q)_{\infty}^{N}}{2^N N!}
\int_{{\mathbb T}^N}
\prod_{1 \leq i < j \leq N} \frac{\Gamma((pq)^{s}
z_i^{\pm 1} z_j^{\pm 1};p,q)} { \Gamma(z_i^{\pm 1} z_j^{\pm 1};p,q) }
    \\    && \makebox[2em]{}   \times
\prod_{j=1}^N \frac{\prod_{i=1}^4
\Gamma((pq)^{r_q}v^{-2} y_i z^{\pm 1}_j;p,q) \prod_{i=5}^8
\Gamma((pq)^{r_{\widetilde q}}v^2 y_i z^{\pm 1}_j;p,q)}
{\Gamma(z_j^{\pm 2};p,q)}\frac{d z_j}{2 \pi i z_j}.
\label{IM1}\end{eqnarray}

The equality $I_E \ = \ I_M^{(1)}$ follows from the Rains transformation
(\ref{Rains1}) after using the relations
\begin{eqnarray} \nonumber
&&\prod_{1\leq j<k\leq 4}
\Gamma\left(\rho^{-2}t_jt_k;p,q,t\right)
= \prod_{1\leq j<k\leq 4}
\Gamma\left(\frac{pq t^{1-N}}{t_1t_2t_3t_4}t_jt_k;p,q,t\right)
\\ \nonumber && \makebox[2em]{}
 = \prod_{1\leq j<k\leq 4}\Gamma\left(\frac{pq
t^{1-N}}{t_jt_k};p,q,t\right)=\prod_{1\leq j<k\leq 4}
\Gamma\left(t^Nt_jt_k;p,q,t\right)
    \\ \nonumber     && \makebox[2em]{}
= \prod_{1\leq j<k\leq 4} \left(\prod_{l=0}^{N-1}
\Gamma\left(t^lt_jt_k;p,q\right)\right)
\Gamma\left(t_jt_k;p,q,t\right),
\nonumber \end{eqnarray}
since
$$
\Gamma(pqt z;p,q,t) \ = \ \Gamma(z^{-1};p,q,t).
$$

\subsection{Second class of dualities}

The second class of dual magnetic theories has the same flavour
group as in the previous case but different representation content.
Again, we expect that there are 35 dual models in this class
whose generic representative is described in Table 9.
\begin{center}
Table 9.
\begin{tabular}{|c|c|c|c|c|c|c|}
  \hline
                & $SP(2N)$            & $SU(4)$    & $SU(4)$    & $U(1)_B$ & $U(1)$ & $U(1)_R$            \\  \hline
  $q$           & $f$                & $\overline{f}$   & 1   & $1$     &   $-\frac{N-1}{4}$    & $\frac{1}{4}$  \\  \hline
 $\widetilde{q}$& $f$   & 1            &$\overline{f}$&
$-1$   &   $-\frac{N-1}{4}$    & $\frac{1}{4}$ \\  \hline
  $Y$             & $T_A$       & 1            &    1         & 0    & 1    & 0                          \\  \hline
  $M_J $          & 1              & $f$            &    $f$         & 0    & $\frac{2J-N+1}{2}$    & $\frac 12$       \\  \hline
  $\widetilde{V}$             & $adj$              & 1            &    1         & 0     & 0    &  $\frac 12$   \\  \hline
\end{tabular}
\end{center}

Similarly to the previous case, we find
\begin{eqnarray} && \makebox[-2em]{}
i_M(p,q,z,\widetilde{y},\widehat{y}) = - \left( \frac{p}{1-p} + \frac{q}{1-q} \right)
\chi_{SP(2N),adj}(z ) \\
\nonumber && \makebox[2em]{}  + \frac{1}{(1-p)(1-q)} \Biggl\{
(pq)^{r_Y} \chi_{SP(2N),T_A}(z) - (pq)^{1-r_Y}
\chi_{SP(2N),\overline{T}_A}(z ) \\
\nonumber && \makebox[2em]{} + (pq)^{r_q}v
  \chi_{SU(4),\overline{f}}(\widetilde{y}) \chi_{SP(2N),f}(z) -
(pq)^{1-r_q}\frac{1}{v} \chi_{SU(4),f}(\widetilde{y})
\chi_{SP(2N),\overline{f}}(z ) \\
\nonumber   && \makebox[2em]{}  +
(pq)^{r_{{\widetilde{q}}}}\frac{1}{v} \chi_{SU(4),\overline{f}}(\widehat{y})
\chi_{SP(2N),f}(z ) - (pq)^{1-r_{\widetilde{q}}} v
\chi_{SU(4),f}(\widehat{y})
\chi_{SP(2N),\overline{f}}(z ) \\
\nonumber && \makebox[2em]{}  + \sum_{J=0}^{N-1}
 \left( (pq)^{r_{M_J}} \chi_{SU(4),f}(\widetilde{y}) \chi_{SU(4),f}(\widehat{y})
- (pq)^{1-r_{M_J}}
\chi_{SU(4),\overline{f}}(\widetilde{y})\chi_{SU(4),\overline{f}}(\widehat{y})\right) \Biggr\},
\end{eqnarray}
where
$$
r_q =r_{\widetilde{q}}= \frac14 - \frac{N-1}{4} s, \quad
 r_Y = s, \quad
r_{M_J} = \frac12 - \frac12 (N-1-2J) s.
$$
Then the index for this magnetic theory is given by
\begin{eqnarray}\nonumber
 &&   I_M^{(2)} =\Gamma((pq)^{ s };p,q)^{N-1} \prod_{J=0}^{N-1}
\prod_{i=1}^4 \prod_{j=5}^8 \Gamma((pq)^{r_{M_J}} y_i y_j;p,q)
    \nonumber \\ \nonumber   && \makebox[2em]{} \times
\frac{(p;p)_{\infty}^{N} (q;q)_{\infty}^{N}}{2^N N!}\int_{{\mathbb T}^N}
\prod_{1 \leq i < j \leq N}
\frac{\Gamma((pq)^{ s } z_i^{\pm 1} z_j^{\pm 1};p,q)}
{\Gamma(z_i^{\pm 1} z_j^{\pm 1};p,q)}
    \\     &&  \makebox[2em]{} \times
 \prod_{j=1}^N \frac{\prod_{i=1}^4
\Gamma((pq)^{r_q} v^2y_i^{-1} z^{\pm 1}_j;p,q) \prod_{i=5}^8
 \Gamma((pq)^{r_{\widetilde q}}v^{-2} y^{-1}_i z^{\pm 1}_j;p,q)}
{\Gamma(z_j^{\pm 2};p,q)}\frac{d z_j}{2 \pi i z_j},
\label{IM2}\end{eqnarray}
where we have chosen the same relations between the characteristic
 variables $v,\widetilde{y}_j, \widehat{y}_j$ and $y_j$ as for $I_M^{(1)}$.
In order to prove $I_E=I_M^{(2)}$, it is necessary to repeat the Rains transformation
twice with the parameters $s_3,s_4,s_5,s_6$ playing the role of $t_1,t_2,t_3,t_4$
in the same way as was done in the $N=1$ rank case.

\subsection{Third dual picture}

Finally, there is only one representative in the third class of
magnetic theories. It has the symmetry groups
$$
{G}\ =\ SP(2N), \qquad {F}\ =\ SU(8) \times U(1),
$$
and its fields content is fixed in Table 10.
\begin{center}
Table 10.
\begin{tabular}{|c|c|c|c|c|}
  \hline
   & $SP(2N)$ & $SU(8)$ & $U(1)$ & $U(1)_R$ \\  \hline
  $q$ & $f$ & $\overline{f}$ & $-\frac{N-1}{4}$ & $\frac{1}{4}$ \\  \hline
  $Y$ & $T_A$ & 1 & 1 & 0  \\ \hline
  $M_J $ & 1 & $T_A$ & $\frac{2J-N+1}{2}$ & $\frac 12$ \\    \hline
  $\widetilde{V}$ & $adj$ & 1 & 0 & $\frac 12$  \\  \hline
\end{tabular}
\end{center}
This dual theory was constructed originally in \cite{Csaki1}.

The single particle state index in this case is
\begin{eqnarray} && \makebox[-2em]{}
i_M(p,q,z,y) = - \left( \frac{p}{1-p} + \frac{q}{1-q} \right)
\chi_{SP(2N),adj}(z ) \\
\nonumber && \makebox[2em]{}  + \frac{1}{(1-p)(1-q)} \Biggl\{
(pq)^{r_Y} \chi_{SP(2N),T_A}(z) - (pq)^{1-r_Y}
\chi_{SP(2N),\overline{T}_A}(z ) \\
\nonumber && \makebox[2em]{} + (pq)^{r_{{q}}}
\chi_{SU(8),\overline{f}}(y) \chi_{SP(2N),f}(z ) - (pq)^{1-r_{q}}
\chi_{SU(8),f}(y)
\chi_{SP(2N),\overline{f}}(z ) \\
\nonumber && \makebox[2em]{}  + \sum_{J=0}^{N-1}
 \left( (pq)^{r_{M_J}} \chi_{SU(8),T_A}(y)
- (pq)^{1-r_{M_J}} \chi_{SU(8),\overline{T}_A}(y) \right) \Biggr\},
\end{eqnarray}
where
$$
r_q \ = \ \frac{1-s(N-1)}{4},\quad r_Y=s,\quad r_{M_J} \ = \
sJ+\frac{1-s(N-1)}{2}.
$$
The magnetic superconformal index has the form
\begin{eqnarray}\nonumber
 &&   I_M^{(3)}= \Gamma((pq)^{r_Y};p,q)^{N -1}
\prod_{J=0}^{N-1} \prod_{1 \leq i < j \leq 8} \Gamma((pq)^{r_{M_J}}
y_i y_j;p,q)
    \\ \nonumber     && \makebox[2em]{} \times
\frac{(p;p)_{\infty}^{N} (q;q)_{\infty}^{N}}{2^N N!}\int_{{\mathbb
T}^N} \prod_{1 \leq i < j \leq N} \frac{\Gamma((pq)^{r_Y} z_i^{\pm
1} z_j^{\pm 1};p,q)} {\Gamma(z_i^{\pm 1} z_j^{\pm 1};p,q)}
    \\   && \makebox[4em]{} \times
\prod_{j=1}^N  \frac{\prod_{i=1}^8
\Gamma((pq)^{r_q} y_i^{-1} z_j^{\pm 1};p,q)}
{ \Gamma(z_j^{\pm 2};p,q)} \frac{d z_j}{2 \pi i z_j},
\label{IM3}\end{eqnarray}
Equality $I_E=I_M^{(3)}$ follows from a triple application of the
key identity (\ref{Rains1}) similar to the $N=1$ case considered
earlier.
For a special quantized value of the parameter $s=\frac{1}{N+1}$,
this result describes equality of superconformal indices in
the Kutasov-Schwimmer dual models with the $SP(2N)$ gauge group,
the number of flavour $N_f=4$, and a special value of the corresponding parameter
$k=N$, see \cite{Kutasov,KS,DO}. After a reduction to the
$s$-confining theory (see below), one obtains equality of indices
for $N_f=3,\, k=N$ case as well.
As to the 't~Hooft anomaly matching conditions
for our new dual models -- we have verified that all of them are satisfied.

\section{Reduction to six flavours}

If we take $t_7t_8=pq$ (or $y_7y_8=(pq)^{1/2}$) for the $SP(2)$-group case,
then, because of the reflection identity $\Gamma(a,b;p,q)=1$ for $ab=pq$,
the integral $I(t_1, \ldots , t_8;p,q)$
is reduced to the left-hand side of (\ref{ell-int}). In physical terms
this means that we add to the $SP(2)$ gauge group SQCD Lagrangian mass terms
for two components of the quark superfields and tend their masses to infinity
washing away them from the spectrum.
As to the integral $I_M^{(1)}$, in this limit two pairs of poles
pinch the contour of integration $\mathbb{T}$ and integral's value
is given by the sum of corresponding residues which yields the right-hand
side expression in (\ref{ell-int}). Physically this means that
the corresponding dual magnetic theory is the Wess-Zumino model of
appropriate meson fields, and the electric theory has confinement.

However, if we set $y_4y_5=(pq)^{1/2}$, then the integral $I_M^{(1)}$
gets simplified, but there is no pinching of the contour and there remains a
nontrivial integral. Physically this means that addition of large mass terms
to different quark superfield components reduces the number of flavours to 6,
but it keeps the gauge group $SP(2)$ intact with the flavour group being
reduced to $SU(3)_l\times SU(3)_r \times U(1)_B\times U(1)_{add}$. Note that
the latter group is of rank $6$ whereas the electric theory has $SU(6)$
flavour group of rank 5.

For the second class dual models the situation is different. For $y_7y_8=(pq)^{1/2}$
there is no pinching of the contour in $I_M^{(2)}$. This integral gets simplified, but
remains a non-trivial integral. The corresponding SQCD model has the
non-trivial gauge group $G=SP(2)$ and
$F=SU(4)\times SU(2)\times SU(2)_{add} \times U(1)_B$. Again, this flavour
group has rank 6. Vice versa, for $y_4y_5=(pq)^{1/2}$
one finds pinching of the contour in $I_M^{(2)}$, the integration disappears, and one
comes to the $s$-confinement with the plain meson fields theory. In the third class
dual model there is only one option -- for any $y_jy_k=(pq)^{1/2}$
the contour in  $I_M^{(3)}$ is pinched, gauge group disappears,
and one comes to the $s$-confinement.

Similar picture holds for $SP(2N),\, N>1,$ gauge group case. Skipping
the details, we present the corresponding non-trivial field theories in
the tables below. The electric theory is described in Table 11.

\begin{center}
 Table 11.
\begin{tabular}{|c|c|c|c|c|}
  \hline
   & $SP(2N)$ & $SU(6)$ & $U(1)$ & $U(1)_R$ \\  \hline
  $Q$ & $f$ & $f$ & $-\frac{N-1}{3}$ & $\frac 16$ \\  \hline
  $X$ & $T_A$ & 1 & 1 & $0$ \\ \hline
  $V$ & $adj$ & 1 & 0 & $\frac 12$  \\  \hline
\end{tabular}
\end{center}

The first class dual models with nontrivial gauge group are described in
Table 12. Equality of the corresponding superconformal indices is obtained
after mere substitution of the constraint $y_4y_5=(pq)^{(1+(N-1)s)/2}$
into formulas (\ref{IE}) and (\ref{IM1}).

\begin{center}
Table 12.
\begin{tabular}{|c|c|c|c|c|c|c|c|}
  \hline
   & $SP(2N)$ & $SU(3)$ & $SU(3)$ & $U(1)$ &$U(1)_B$ & $U(1)_{add}$
& $U(1)_R$ \\  \hline
  $q$ & $f$ & $f$ & 1 & $-\frac{N-1}{3}$ & -1 & -1 & $\frac{1}{6}$ \\  \hline
  $\widetilde{q}$ & $f$ & 1 & $f$ & $-\frac{N-1}{3}$ & 1 & 1 &
$\frac{1}{6}$ \\  \hline
  $M_{1J}$ & 1 & $T_A=\overline{f}$ & 1 & $J-2\frac{N-1}{3}$ & 4 & 0 &
$\frac 13$  \\ \hline
  $N_{1J}$ & 1 & $f$ & 1 & $J-\frac{N-1}{3}$ & 2 & 2 & $\frac 23$  \\ \hline
  $M_{2J}$ & 1 & 1 & $T_A=\overline{f}$ & $J-2\frac{N-1}{3}$ & -4 & 0 &
$\frac 13$  \\ \hline
  $N_{2J}$ & 1 & 1 & $f$ & $J-\frac{N-1}{3}$ & -2 & -2 & $\frac 23$  \\
\hline
  $Y$ & $T_A$ & 1 & 1 & 1 & 0 & 0 & $0$ \\ \hline
  $\widetilde{V}$ & $adj$ & 1 & 1 & 0 & 0 & 0 & $\frac 12$  \\  \hline
\end{tabular}
\end{center}

The second class dual models with the nontrivial gauge group are described in
Table 13. Equality of the corresponding indices is obtained
after substitution of the constraint $y_7y_8=(pq)^{(1+(N-1)s)/2}$ into formulas
(\ref{IE}) and (\ref{IM2}).

\begin{center}
Table 13.
\begin{tabular}{|c|c|c|c|c|c|c|c|}
  \hline
   & $SP(2N)$ & $SU(4)$ & $SU(2)_{add}$ & $SU(2)$ & $U(1)$ &
$U(1)_B$ & $U(1)_R$ \\  \hline
  $q$ & $f$ & $\overline{f}$ & 1 & 1 & $-\frac{N-1}{3}$ & -1 &
$\frac{1}{6}$ \\  \hline
  $\tilde q$ & $f$ & 1 & $f$ & 1 & $-\frac{N-1}{3}$ & 2 & $\frac{1}{6}$ \\  \hline
  $M_J$ & 1 & $f$ & $f$ & 1 & $J-\frac{N-1}{3}$ & -1 & $\frac 23$  \\ \hline
  $N_J$ & 1 & $f$ & 1 & $f$ & $J-2\frac{N-1}{3}$ & 1 & $\frac 13$  \\ \hline
  $Y$ & $T_A$ & 1 & 1 & 1 & 0 & 0 & $0$ \\ \hline
  $\widetilde{V}$ & $adj$ & 1 & 1 & 0 & 0 & 0 & $\frac 12$  \\  \hline
\end{tabular}
\end{center}

Finally, the field content of the model without
gauge group is fixed in Table 14, where $k=2,\ldots,N$ and $J=0,\ldots,N-1$.
\begin{center}
Table 14.
\begin{tabular}{|c|c|c|c|}
  \hline
    & $SU(6)$ & $U(1)$ & $U(1)_R$ \\  \hline
  $M_k$ & 1 & $k$ & 0 \\  \hline
  $N_J$ & $T_A$ & $J-2\frac{N-1}{3}$ & $\frac 13$ \\  \hline
\end{tabular}
\end{center}
For completeness, we present explicitly  equality of superconformal indices
for this case in the appropriate notation:
\begin{eqnarray}\nonumber
&& I_E=\frac{(p;p)_\infty^N (q;q)_\infty^N}{2^NN!}
\Gamma(t;p,q)^{N-1}
\int_{\mathbb{T}^N} \prod_{1\leq j<k\leq n}
\frac{\Gamma(tz_j^{\pm 1} z_k^{\pm 1};p,q)}{\Gamma(z_j^{\pm 1} z_k^{\pm 1};p,q)}
\\ \nonumber && \makebox[8em]{} \times
\prod_{j=1}^N\frac{\prod_{m=1}^6\Gamma(t_mz_j^{\pm 1};p,q)}{\Gamma(z_j^{\pm2};p,q)}
\frac{dz_j}{2\pi iz_j}
\\ && \makebox[2em]{}
=I_M= \prod_{j=2}^N\Gamma(t^j;p,q)\prod_{J=0}^{N-1}
\prod_{1\leq k<m\leq 6}\Gamma(t^Jt_kt_m;p,q),
\label{e-selberg}\end{eqnarray}
where $|p|, |q|,$ $|t|,$ $|t_m| <1,$ and $t^{2n-2}\prod_{m=1}^6t_m=pq$.
This relation describes the elliptic analogue of the
Selberg integral for the $BC_N$-root system \cite{Spiridonov1}.
The dual theories of Tables 12 and 13 are new, and the  $s$-confined
model of Table 14 was described in \cite{Csaki3}.

Let us discuss now the possible number of $N_f=6$ dual models. To count them one
has to describe the group structure of integrals remaining after imposing the
constraint $t_7t_8=pq$. In the notation used for the description of $W(E_7)$
in the end of Sect. 3.1, it is equivalent to the constraint $x_7+x_8=const.$
This reduces the $E_7$ root system to $E_6$. The Weyl group $W(E_6)$ includes
the evident $S_6\times S_2$ group permuting first six and last two coordinates
of $x=(x_1,\ldots,x_6; x_7,x_8)$ between themselves. It is generated by
the $R_v$-reflections for the vectors $v\in \pm (e_i-e_j)$ for
$1\leq i<j\leq 6$ or $i=7,\, j=8$. Other nontrivial 20 vectors are obtained
by the $R_v$-reflections of the $S_6\times S_2$ orbit of $x$ for the vectors
\begin{equation}
v\in \frac12 \left( \sum_{k=1}^8 (-1)^{\mu_k} e_k \right), \quad
\mu_k\in\{0,1\},\quad \sum_{k=1}^6\mu_k=3,\quad \mu_7+\mu_8=1
\label{E6-vectors}\end{equation}
leading to the coordinate tranformations
$$
x_j'=x_j  -\frac14 \sum_{k=1}^8 (-1)^{\mu_j+\mu_k} x_k,
$$
where $j=1,\ldots,8$.
The remaining 15 nontrivial vectors of the $W(E_6)$-orbit are obtained by the
reflections $R_vR_{v'}$ with $v,\,v'$ from (\ref{E6-vectors}). They have
coordinates of the form
$$
x_{k_1,k_2,k_3,k_4}'=-x_{k_1,k_2,k_3,k_4}+\frac12 (x_{k_1}+x_{k_2}+x_{k_3}+x_{k_4}),
$$
$$
x_{k_5}'=-x_{7}+\frac12 (x_{7}+x_{8}+x_{k_5}+x_{k_6}),\quad
x_{k_6}'=-x_{8}+\frac12 (x_{7}+x_{8}+x_{k_5}+x_{k_6}),
$$
where $k_1,\ldots,k_6\in \{1,\ldots,6\}$ for $k_i\neq k_j$ and, finally,
$$
x_{7}'=-x_{k_5}+\frac12 (x_{7}+x_{8}+x_{k_5}+x_{k_6}),\quad
x_{8}'=-x_{k_6}+\frac12 (x_{7}+x_{8}+x_{k_5}+x_{k_6}).
$$
Since $\dim \{W(E_6)/(S_6\times S_2)\}=36$, it is expected that
there are 36 dual models with the nontrivial gauge group $G=SP(2N)$ and $N_f=6$.
The rest of 36 dual models with $N_f=8$ reduce for $N_f=6$ to one additional
37th $s$-confined dual model without gauge group (which becomes completely Higgsed).

There is an interesting problem of comparing anomalies for $N_f=6$ theories.
It is not difficult to check validity of 't~Hooft's criterion for
the electric and confined theories.
However, the first and second class dual models have rather different flavour groups
explicitly seen in UV. To compare with the electric theory, one can
check first that all anomalies associated with $SU(2)_{add}$ and $U(1)_{add}$
groups vanish. Then it is necessary to embed the remaining parts of the
magnetic flavour groups into $SU(6)\times U(1)$ and match the corresponding
anomalies in the standard way. The missing anomalies for the cosets
$SU(6)/(SU(3)\times SU(3)\times U(1)_B)$
and $SU(6)/(SU(4)\times SU(2)\times U(1)_B)$ may, probably,
be imitated by some nonlinear chiral models added to the corresponding SQCD's.
If we compare anomalies of the first and second class dual magnetic models between
themselves, it is necessary to go further and split both flavour groups without
$SU(2)_{add}$ and $U(1)_{add}$ pieces to the
smaller subgroup $SU(3)\times SU(2)\times U(1)_1 \times U(1)\times U(1)_B$,
for which the anomalies match in the standard way. The rest of the
anomalies for non-explicit pieces of the flavour symmetries may, probably, be
related to (unknown) non-linear chiral models incorporated into
both magnetic theories.

\section{Conclusion}

To conclude, in this paper we have used known $W(E_7)$-group
transformation identities for elliptic
hypergeometric integrals in order to describe some known
and new Seiberg dualities for ${\mathcal N}=1$ supersymmetric field theories
with $SP(2N)$ gauge groups and the number of flavours $N_f=8$ and $N_f=6$.
We expect that there are 72 self-dual theories for $N_f=8$, among which
only four have essentially different field content and symmetry groups.
For $N_f=6$ we expect existence of 36 dual theories with the non-trivial gauge group
(with only three essentially different field content models)
and one $s$-confined meson fields theory.
 The flavour groups for $N=1$ and $N>1$
differ from each other. The tables for $N=1$ can be obtained from those of $N>1$
after setting $J=0,\; N=1$ and deleting one row and one column. We decided
to give separate consideration of the $N=1$ case because
{\em all} superconformal indices for dual theories known to us involve
generalizations of one or another
transformation of the corresponding electric theory characteristic variables.
For instance, there is an interesting reduced form of the multiple duality phenomenon
for $G=SU(N)$ gauge groups for $N>2$ \cite{Csaki2,spi-var:elliptic}.

It turns out that the connection of superconformal indices with the elliptic
hypergeometric integrals leads to some new results in the theory of
elliptic hypergeometric functions. Namely, there are
new conjectures for both -- the elliptic beta integrals and transformation
identities for higher order elliptic hypergeometric functions on
root systems. For example, there is an almost complete match of the
list of $s$-confining theories in \cite{Csaki3} and elliptic beta
integrals on root systems listed in \cite{Spiridonov1},
with one of the known integrals leading to a new example of
$s$-confining theories \cite{spi-var:elliptic}.
Vice versa, e.g., an analysis of the $s$-confining duality for the exceptional
gauge group $G_2$ of \cite{Exceptional} leads to the following new
elliptic beta integral
\begin{eqnarray} \nonumber
&&
\frac{(p;p)_\infty^2 (q;q)_\infty^2}{2^23}\int_{{\mathbb T}^2} \frac{\prod_{k=1}^3\prod_{m=1}^5\Gamma(t_mz_k^{\pm1};p,q)}
{\prod_{1\leq j<k\leq3} \Gamma(z_j^{\pm1}z_k^{\pm1};p,q)}
\prod_{k=1}^2\frac{dz_k}{2\pi iz_k}
\\ &&  \makebox[4em]{}
=\prod_{m=1}^5\frac{\Gamma(t_m^2;p,q)}
{\Gamma(t_m;p,q)\Gamma((pq)^{1/2}t_m;p,q)}
\prod_{1\leq l<m\leq 5}
\frac{\Gamma(t_lt_m;p,q)}{\Gamma((pq)^{1/2}t_lt_m;p,q)},
\label{G2}\end{eqnarray}
where $z_1z_2z_3=1$, $|t_m|<1$, and $\prod_{m=1}^5 t_m=(pq)^{1/2}$.
As we have known from a private communication, this formula was
conjectured also earlier by M. Ito.
At the moment, no proof of this relation is known to the authors.

The considerations of \cite{Romelsberger,Kinney, DO} justify
the superconformal index building algorithm only for
marginally deformed free theories  (we are indebted to F. Dolan and
Yu. Nakayama for stressing to us this point). However, we apply it to the
interacting theories and, by some deep reason, it works for
them rather well. Therefore it is necessary to find a more rigorous
derivation of formula (\ref{Ind}) for Seiberg dual theories.

Consider now the constraints on the parameters of our models
coming from the renormalization group analysis.
The original Seiberg duality \cite{Seiberg} is based on the gauge
groups $G_E=SU(N)$ and $G_M=SU(N_f-N)$ with the flavour group
$SU(N_f)_l\times SU(N_f)_r\times U(1)_B$ (we used above different counting of the
number of flavours which corresponds to $2N_f=6,8$ in the Seiberg notation).
Existence of the asymptotic freedom in the electric theory leads to
the constraint $N_f < 3N$. Similar requirement for the magnetic theory
yields the bound $3N/2 < N_f$. The combination of two restrictions is called
the conformal window. Formally, for $N=2$ and $N_f=4$
the corresponding models lie in the conformal window.
However, for all our theories the lower bound $3N/2 < N_f$ is not relevant.
In the context of $SP(2N)\leftrightarrow SP(2(N_f-N-2))$
duality with $SU(2N_f)$ flavour groups found in \cite{IP}, the conformal
window has the form $3(N+1)/2< N_f < 3(N+1)$. Formally, for $N=1$ and $N_f=4$ we
have again a pair of models satisfying this constraint, but the lower bound
of this window is not relevant again. The reason for the absence of lower bounds
stems from the self-duality of our models. Indeed, for all of them the rank of
the dual gauge group is fixed, and it does not depend on the number of flavours.
As a result, all our $G=SP(2N)$ models are simultaneously automatically
asymptotically free (both, for $N=1$ and $N>1$). If we consider these models
with arbitrary number of flavours $2N_f$ in the fundamental representation,
then for all of them the one loop beta function is
$\beta(g)=-g^3 (2N+4-N_f)/8\pi^2$. Asymptotic freedom
is guaranteed by the universal bound $N_f<2N+4$, which
is satisfied in our case $N_f=4$ for arbitrary $N\geq 1$.
Let us remark also that at the infrared fixed point, the dimensions of
gauge-invariant scalar fields $\Delta$ are determined by the $R$-charges,
$\Delta=3R/2$. All our meson fields have thus the dimensions 3/2
satisfying the unitarity constraints $\Delta\geq 1$.

As to the 't Hooft anomaly matching conditions -- they are satisfied
pairwise for all dual theories described above for smaller flavour groups.
It looks like that the
key properties needed for this matching are encoded into the balancing conditions
and the $SL_{\tau}(2;{\mathbb Z})$ or $SL_{\sigma}(2;{\mathbb Z})$
 modular group invariance of ``totally" elliptic functions hidden in
the structure of superconformal indices \cite{Spiridonov1}.
(Here the modular variables $\tau $ and $\sigma$ are related to $p$ and $q$ as
$p=e^{2\pi i\tau}$ and $q=e^{2\pi i\sigma}$.)

As a final remark, we would like to speculate on the
relevance of the exceptional root system $E_7$.
The well known Kramers-Wannier duality relates 2D Ising models for low
and high temperatures. Existence of the unifying model with the
Hamiltonian allowing for an explicit transformation of relevant
degrees of freedom makes this duality easy to understand.
Putting Seiberg duality in a similar context, it looks like that
the global symmetry group of the ``brane" (higher-dimensional)
theory unifying all the Seiberg dual theories for $SP(2N)$ groups
is $E_7$, and it is different ``degenerations" that
lead to either $SU(8)\times U(1)$ or $SU(4)_l\times SU(4)_r\times U(1)_B\times U(1)$
flavour groups. In any case, the brane dynamics reproducing
Seiberg duality for $SU(2)$ gauge group is expected to be
more complicated than that described in \cite{GK}.
In order to clarify the origins of this picture
it is necessary to build superpotentials and nonlinear chiral models for our
dualities distinguishing them from each other (like in the triality of \cite{IS}) and
to find their place within the AdS/CFT correspondence framework.

\section*{Acknowledgments}
V.S. is indebted to A.M. Povolotsky for drawing attention
to paper \cite{DO} and to V.A. Rubakov for many helpful advises.
G.V. would like to thank D.I. Kazakov and A.F. Oskin for valuable discussions.
We are grateful to F.A. Dolan and H. Osborn for detailed remarks
on the first version of this paper. In particular, we thank H. Osborn
for a question on the structure of the coset $W(E_7)/S_8$ which
helped us to count the number of new dualities.  We thank also
A. Kmelnitsky for clarifying flavour group structures for some
of the dual models and M. Shifman for the indication on
a possible way to resolve the anomaly matching puzzle.
The organizers of IV-th Sakharov conference on physics
(Moscow, May 2009), Conformal field theory workshop (Chernogolovka, June 2009),
and XVI-th International congress on mathematical physics (Prague, August 2009)
are thanked for giving to us an opportunity to present the results of our work
at these meetings.

The work of V.S. is partially supported by RFBR grant no. 09-01-00271.
The work of G.V. is partially supported by the Dynasty foundation, RFBR
grant no. 08-02-00856 and grant of the Ministry of Education and
Science of the Russian Federation no. 1027.2008.2.

\section{Appendix. Characters for unitary and symplectic groups}

A character $\chi_R(g)$ for $g \in SU(N)$ is a function of the
complex eigenvalues of $g$
$$
x=(x_1, \ldots , x_N), \quad  \prod_{i=1}^Nx_i=1.
$$
The characters of the fundamental and antifundamental representations
of $SU(N)$ group are given by
\begin{eqnarray*} &&
\chi_{SU(N),f}(x) \ = \ \sum_{i=1}^N x_i, \quad
\chi_{SU(N),\overline{f}}(x) \ = \ \chi_{SU(N),f}(x^{-1}).
\end{eqnarray*}
We use also general properties of the characters
\begin{eqnarray*} &&
\chi_{f_1\oplus f_2}=\chi_{f_1}+\chi_{f_2},\quad
\chi_{f_1\otimes f_2}=\chi_{f_1}\chi_{f_2}.
\end{eqnarray*}

For the adjoint representation one has
$\chi_{SU(N),adj}(x) = (\sum_{i=1}^N x_i)(\sum_{j=1}^Nx_j^{-1})-1.$
The character for the anti-symmetric tensor representation of
$SU(N)$ is
$$
\chi_{SU(N),T_A}(x) \ = \ \sum_{1 \leq i < j \leq N} x_i x_j,\quad
 \chi_{SU(N),\overline{T}_A}(x) \ = \ \chi_{SU(N),T_A}(x^{-1}).
$$

A character $\chi_R(g)$ for $g \in SP(2N)$ is a function of the
complex eigenvalues of $g$, $x=(x_1, \ldots , x_N).$ The characters of
the fundamental and antifundamental representations of $SP(2N)$
group have the form
$$
\chi_{SP(2N),f}(x) \ = \chi_{SP(2N),\overline{f}}(x)=\
\sum_{i=1}^N (x_i + x_i^{-1}).
$$
The character for the adjoint representation of $SP(2N)$ is
$$
\chi_{SP(2N),adj}(x) \ = \ \sum_{1 \leq i < j \leq N} (x_ix_j +
x_ix_j^{-1} + x_i^{-1}x_j + x_i^{-1}x_j^{-1}) +
\sum_{i=1}^N(x_i^2+x_i^{-2}) + N.
$$
For $N=1$ it coincides with the adjoint representation character
for $SU(2)$ group.
The character for the anti-symmetric tensor representation of
$SP(2N)$ is
$$
\chi_{SP(2N),T_A}(x) \ = \ \sum_{1 \leq i < j \leq N} (x_ix_j +
x_ix_j^{-1} + x_i^{-1}x_j + x_i^{-1}x_j^{-1}) + N-1.
$$


\begin{thebibliography}{99}

\bibitem{Spiridonov1} V. P. Spiridonov, \textit{Essays on the theory of
elliptic hypergeometric functions}, Uspekhi Mat. Nauk {\bf 63}
 (2008) no. 3, 3--72 (Russian Math. Surveys {\bf 63} (2008) 405--472),
arXiv:0805.3135 [math.CA].

\bibitem{Romelsberger} C. R\"omelsberger, \textit{Counting chiral
primaries in ${\mathcal N}=1$, $d=4$ superconformal field theories},
Nucl. Phys. {\bf B747} (2006) 329--353, hep-th/0510060; \textit{Calculating
the Superconformal Index and Seiberg Duality}, arXiv:0707.3702
[hep-th].

\bibitem{Kinney} J. Kinney, J. M. Maldacena, S. Minwalla and S.
Raju,  \textit{An Index for 4 dimensional super conformal theories},
Commun. Math. Phys. {\bf 275} (2007) 209--254, hep-th/0510251.

\bibitem{S02} N. Seiberg, \textit{Exact results on the space
of vacua of four-dimensional SUSY gauge
theories}, Phys. Rev. {\bf D49} (1994) 6857--6863, hep-th/9402044.

\bibitem{Seiberg} N. Seiberg,  \textit{Electric--magnetic duality in
supersymmetric non-Abelian gauge theories}, Nucl. Phys. {\bf B435}
(1995) 129--146, hep-th/9411149.

\bibitem{DO} F. A. Dolan and H. Osborn,  \textit{Applications of
the Superconformal Index for Protected Operators and $q$-Hypergeometric
Identities to ${\mathcal N}=1$ Dual Theories}, Nucl. Phys. {\bf B818}
(2009) 137--178, arXiv:0801.4947 [hep-th].

\bibitem{Spiridonov2} V. P. Spiridonov,
{\em On the elliptic beta function},
Uspekhi Mat. Nauk {\bf 56} (2001) no. 1, 181--182
(Russian Math. Surveys {\bf 56} (2001) 185--186).

\bibitem{Kutasov} D. Kutasov,  \textit{A comment on duality in ${\mathcal N}=1$
supersymmetric non-Abelian gauge theories}, Phys. Lett. {\bf B351}
(1995) 230--234, hep-th/9503086.

\bibitem{KS} D. Kutasov and A. Schwimmer,
\textit{On duality in supersymmetric Yang-Mills
theory}, Phys. Lett. {\bf B354} (1995) 315--321, hep-th/9505004.

\bibitem{IS} K. Intriligator and N. Seiberg,  \textit{Duality, Monopoles,
Dyons, Confinement and Oblique Confinement in Supersymmetric $SO(N_c)$ Gauge
Theories}, Nucl. Phys. {\bf B444} (1995) 125--160, hep-th/9503179.

\bibitem{I} K. Intriligator,  \textit{New RG fixed points
and duality in supersymmetric $SP(N_c)$ and $SO(N_c)$ gauge
theories}, Nucl. Phys. {\bf B448} (1995) 187--198, hep-th/9505051.

\bibitem{IP}
K. Intriligator and P. Pouliot,  \textit{Exact superpotentials, quantum
vacua and duality in supersymmetric $SP(N_c)$ gauge theories},
Phys. Lett. {\bf B353} (1995) 471--476, hep-th/9505006.

\bibitem{Exceptional} I. Pesando, \textit{ Exact results for the
supersymmetric $G_2$ gauge theories}, Mod. Phys. Lett. {\bf A10} (1995) 1871--1886,
hep-th/9506139; \\ S. B. Giddings and J. M. Pierre, \textit{ Some
exact results in supersymmetric theories based on exceptional
groups}, Phys. Rev. {\bf D52} (1995) 6065--6073, hep-th/9506196.

\bibitem{Csaki1} C. Cs\'aki, W. Skiba and M. Schmaltz,
\textit{Exact results and duality for $SP(2N)$ SUSY gauge theories with
an antisymmetric tensor}, Nucl. Phys. {\bf B487} (1997) 128--140, hep-th/9607210;

\bibitem{Csaki2} C. Cs\'aki, M. Schmaltz, W. Skiba and J. Terning,
\textit{Selfdual ${\mathcal N}=1$
SUSY gauge theories}, Phys. Rev. {\bf D56} (1997) 1228--1238,
hep-th/9701191.

\bibitem{Csaki3} C. Cs\'aki, M. Schmaltz and W. Skiba,
\textit{Confinement in ${\mathcal N}=1$ SUSY Gauge
Theories and Model Building Tools}, Phys. Rev. {\bf D55} (1997) 7840--7858,
hep-th/9612207.

\bibitem{Spiridonov3} V. P. Spiridonov, \textit{Theta hypergeometric
integrals}, Algebra i Analiz {\bf 15} (2003) no. 6, 161--215 (St.
Petersburg Math. J. {\bf 15} (2003) 929--967), math.CA/0303205.

\bibitem{Rains} E. M. Rains, \textit{Transformations of elliptic hypergeometric
integrals},  Ann. of Math., to appear,   math.QA/0309252v4.

\bibitem{spi-var:elliptic} V. P. Spiridonov and  G. S. Vartanov,
\textit{Elliptic hypergeometry of sypersymmetric dualities}, to appear.

\bibitem{Sun} B. Sundborg, \textit{The Hagedorn transition,
Deconfinement and ${\mathcal N}=4$ SYM Theory}, Nucl. Phys.
{\bf B573} (2000) 349--363, hep-th/9908001.

\bibitem{Naka} Yu. Nakayama, \textit{Index for Orbifold Quiver Gauge Theories},
Phys. Lett. {\bf B636} (2006) 132--136, hep-th/0512280.

\bibitem{Plethystic} S. Benvenuti, B. Feng, A. Hanany, and Y. H. He,
\textit{Counting BPS Operators in Gauge Theories: Quivers,
Syzygies and Plethystics}, JHEP {\bf 0711} (2007) 050,
hep-th/0608050;
\\ B. Feng, A. Hanany and Y. H. He,
\textit{Counting gauge invariants: The Plethystic
program}, JHEP {\bf 0703} (2007) 090, hep-th/0701063.

\bibitem{D2} F. A. Dolan, {\em Counting BPS operators in $\mathcal{N}=4$ SYM},
Nucl. Phys. {\bf B790} (2008) 432--464, arXiv:0704.1038 [hep-th].

\bibitem{LS} R. G. Leigh and M. S. Strassler, {\em Accidental symmetries
and $\mathcal{N}=1$ duality in supersymmetric gauge theory},
Nucl. Phys. {\bf B496} (1997) 132--148, hep-th/9611020.

\bibitem{GK} A. Giveon and D. Kutasov, {\em Brane Dynamics and Gauge Theory},
Rev. Mod. Phys. {\bf 71} (1999) 983--1084, hep-th/9802067.

\end{thebibliography}
\end{document}